\documentclass[aps,twocolumn]{revtex4}
\usepackage{graphicx}

\begin{document}

\title{The Effect of Side Traps on Ballistic Transistor in Kondo Regime}
\author{Tetsufumi Tanamoto, Ken Uchida  and Shinobu~Fujita}
\affiliation{Advanced LSI laboratory, Corporate R\&D Center,
Toshiba Corporation,\\ 1 Komukai Toshiba-cho, Saiwai-ku,
Kawasaki 212-8582, Japan
}


\begin{abstract}
 The effect of side traps on current and conductance in 
ballistic transport is calculated 
 using slave-boson mean field theory, particularly 
 when there are electrodes on both sides of a short channel. 
 The depth of the conductance dip, which is due 
 to destructive interference known as the Fano-Kondo effect, 
 depends on the tunneling coupling between 
 the conducting region and the electrodes.
 The results imply that ballistic devices  
 are sensitive to trap sites.
\end{abstract}

\maketitle
\sloppy

\section{Introduction}
As the size of  Si metal-oxide-semiconductor field-effect transistors (MOSFET)
decrease, the electronic transport of carriers 
is expected to change from the drift-diffusive region 
to the ballistic region\cite{Natori1,Natori2,Lundstrom}.
In this era, SiO${}_2$ gate insulators are being 
replaced by higher-dielectric-constant materials 
(high-k materials), 
in which trap states cannot be avoided. 
Trap sites degrade device performance such as by causing 
flat band voltage shifts. 
Thus, the effect of trap sites in gate insulators on transport 
properties is one of the 
important topics of ballistic transistors.

On the other hand, the effect of side-trap states on 
an infinite quantum wire (QW) has been treated as a Fano-Kondo (FK) problem, 
in which side-trap states are constructed by a quantum dot (QD). 
This has attracted great interest, because conductance is suppressed 
as a result of destructive interference at $T\!<\!T_K$ (Kondo temperature) 
\cite{Kondo,Kang,Aligia,Sato}. 
The simplest analytical form of zero-temperature conductance $G$ 
is described by the average number of carriers in the dot 
$\langle n_d\rangle$ 
as $G\!=\!(2e^2/h) \cos^2 (\pi\langle n_d \rangle/2) $, which is in 
contrast with that of an embedded quantum dot (conventional Kondo effect),
$G\!=\!(2e^2/h) \sin^2 (\pi \langle n_d \rangle/2)$, and 
shows a dip when charge is localized in the side QD 
($\langle n_d\rangle \approx 1)$. 

However, an infinite QW without source and drain 
is not representative of future ballistic transistors, 
because the channel length of ballistic transistors 
is sufficiently short. Thus, we cannot directly use the results 
obtained from previous works, and the effect of coupling between
the channel and the electrodes should be detailed.
Here, we investigate the effect of trap sites on a ballistic 
transistor using the Keldysh Green's function method 
based on slave-boson mean field theory (SBMFT)\cite{Newns,Coleman,Tanamoto}. 
\begin{figure}
\includegraphics[width=7cm]{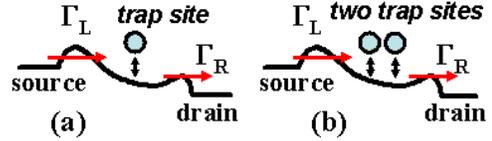}
\caption{Side-trap states near the conducting channel. 
(a) One-trap-site case. (b) Two-trap-site case. 
Potential barriers are assumed to exist between the electrodes 
and the channel region. 
}
\label{Fig.1}
\end{figure}

\section{Formulation}
We model a ballistic transistor, as shown in Fig. 1. 
We assume that two potential barriers exist between the electrodes 
(source and drain) and the ballistically conducting channel.
The tunneling rates between the electrode part and the channel part 
are written as $\Gamma_L$ and $\Gamma_R$, 
respectively. 
This model can also be used for Schottky transistors\cite{Kinoshita1,Kinoshita2}.
Because SBMFT is numerically simple and efficient for the analysis of 
strongly correlated QD systems, this method is 
widely used for the study of the Kondo effect. 
In SBMFT, an infinite on-site Coulomb interaction for each trap site 
is assumed, 
which means that at most one excess electron is permitted in each trap 
site\cite{Newns,Coleman,Tanamoto}.

\subsection{One trap site}
First, we consider the effect of one trap site [Fig.~1(a)]. 
The Hamiltonian is written in terms of slave-boson mean fields 
as $H=H_{\rm chan}^{\rm (I)}+H_{\rm elec}+H_{\rm tran}$. 
$H_{\rm chan}^{\rm (I)}$, $H_{\rm elec}$, and 
$H_{\rm tran}$ represent the conducting channel with the trap site, 
the two electrodes, and the transference of electrons between 
the channel, and the electrodes, respectively: 
%
\begin{eqnarray}
H_{\rm chan}^{\rm (I)}\!&=&\!\!
\sum_{s}
\left\{\sum_{k}
E_{k}c_{ks}^\dagger c_{ks}+\epsilon_f d_s^\dagger d_s  
\right.\nonumber \\
&\!+\!&\left. \!\sqrt{z}\sum_{k}
 [V_d d_{s}^\dagger c_{ks}\!+\!{\rm h.c}]
\right\}
\!+\!(\epsilon_f\!-\!E_D)(z-1)
\\
H_{\rm elec}&=&\sum_{\alpha=L,R}
\sum_{ks}E_{k\alpha}f_{ks}^{\alpha\dagger} f_{ks}^\alpha
\\
H_{\rm tran}&=&\!\!
\sum_{\alpha=L,R}\sum_{k_1k_2s}(t_{k_1k_2}^\alpha c_{k_1s}^\dagger 
f_{k_2s}^\alpha
\!+\!{\rm h.c.}).
\label{Hamiltonian}
\end{eqnarray}
where $f_{ks}^\alpha (\alpha=L,R)$, $c_{ks}$, and $d_s$
are respectively the annihilation electron operator for
both electrodes, the channel region, and the trap site. 
$s$ shows spin components of $s=\uparrow,\downarrow$.
$E_{k\alpha}$ and $E_D$ are the energies of the
electrodes and trap site, respectively. 
Here, for simplicity, we treat a purely one-dimensional system, 
which means that $k$-integrals are carried out only in 
one direction.
$\epsilon_f$ is the quasi-particle trap energy. 
$z$ is the mean value of the boson operator, showing the 
average vacancy rate in the trap site. 
$\epsilon_f$ and $z$ are determined by self-consistent 
equations shown below. 
$t_{k_1k_2}^\alpha$ is the tunneling matrix between 
the channel region and the electrodes, and $V_d$ is that between
the conducting region and the trap site.
We take a constant value for $V_d$, assuming that the tunneling 
barrier to the trap site is sufficiently high.
SBMFT is valid below 
$T_K=D \exp( E_D /(2 V_d^2 N_c(E_F) )$
($D$ is the band width and $N_c(E_F)$ is the density of 
states (DOS) in the channel region at $E_F$) 
\cite{Newns,Coleman,Tanamoto}.

The current $I_D$ between the source and the drain 
is described by the Keldysh Green's function
as 
\begin{equation}
I_D=\frac{2e}{h}\sum_{kk'} \int d\omega
{\rm Re} \left\{ t_{kk'}^L G_{c_{k'}f_{k}^L}^< (\omega) \right\}
\end{equation}
where 
$
G_{c_{k'}f_{k}^L}^< (t,t')$ $\equiv$ $ i\langle f_{k}^{L\dagger}(t') c_{k'}(t)
\rangle
$
\cite{Meir,Datta} (we neglect spin dependence). 
Using the relation $G^<(\omega)=g_1^r(\omega) g_2^<(\omega)
+g_1^<(\omega) g_2^a(\omega)$ 
where $G(\omega)=g_1(\omega)g_2(\omega)$ 
($g_1^r(\omega)$ is the retarded Green's function for $g_1(\omega)$ 
and $g_2^a(\omega)$ is the advanced Green's function for $g_2(\omega)$), 
we can describe $G_{c_{k'}f_{k}^L}^< (t,t')$ 
using elementary Green's functions.
First, the current without any traps is derived as
$I_0=g_0 V_D$ ($V_D$ is the drain voltage), where
\begin{equation}
g_0=\frac{e}{h} \frac{y_0}{(1+y_0)^2} 
\label{g0}
\frac{\Gamma_L \Gamma_R}{\gamma}.
\end{equation}
$y_0\equiv \pi N_c(E_F)\gamma$ is the number of channel electrons 
in the energy width of $\gamma$ 
[$\gamma=(\Gamma_L+\Gamma_R)/2$].
Note that the energy dispersion $E_{k\alpha}$ in the channel region has  
continuum $k$ dependence. This is in contrast with that of a quantum 
dot discussed in refs.\cite{Meir} and \cite{Datta}, where the band mixing of 
discrete energy levels in the quantum dot can be neglected. 
$I_D$ with a trap site is given as
\begin{equation}
I_D =g_0 \int_{-D}^{D} d\omega 
\frac{(\omega-\epsilon_f)^2}{(\omega-\epsilon_f)^2+z^2\eta^2}
(f_L(\omega)-f_R(\omega))
\end{equation}
where $\eta=\eta_0 y_0/(1+y_0)$ with 
$\eta_0=V_d^2 /\gamma$. 
$f_L(\omega) \equiv (\exp((\omega\!-\!E_F\!+eV)/T)\!+1)^{-1}$ and 
$f_R(\omega)\equiv (\exp((\omega\!-\!E_F)/T)\!+1)^{-1}$ are the 
Fermi distribution functions of the left and right electrodes, respectively  
(Boltzmann's constant $k_B=1$).
This formula is the main result of this study and shows that 
the existence of a trap site decreases $I_D$ greatly when 
the energy of carrier electrons is close to the trap site energy. 
Compared with the infinite wire case\cite{Kang,Aligia}, 
we can see that the coupling  strength $\eta$ 
is modified by $y_0$ and a function of $\Gamma_L$ and $\Gamma_R$.

The self-consistent equations for $\epsilon_f$ and $z$ are given as
\begin{eqnarray}
2\!\int_{-D}^D \frac{d\omega}{\pi}\!\!\frac{\eta (\omega-\epsilon_f)}
{(\omega-\epsilon_f)^2+z^2 \eta^2} F_1(\omega)
&=&
E_D\!-\!\epsilon_f
\\
2\!\int_{-D}^D \frac{d\omega}{\pi}
\frac{z\eta}
{(\omega-\epsilon_f)^2+z^2 \eta^2} F_1(\omega)
\!&=&\!1\!-\!z 
\label{Self}
\end{eqnarray}
where 
$F_1(\omega) \equiv 
\{y_0 [\Gamma_L f_L(\omega)\!+\!\Gamma_R f_R(\omega))]
/(\Gamma_L\!+\!\Gamma_R)
\!+\! f_c(\omega)\}/(1\!+\!y_0)$ and 
$f_c(\omega) \equiv (\exp((\omega-E_F\!+\!eV/2)/T)\!+\!1)^{-1}$. 
In the $\gamma \rightarrow 0$ limit, these equations 
reduce to those given in refs.~\cite{Newns} and \cite{Coleman}.
As shown below, the $V_D$ dependence of $\epsilon_f$ and 
$z$ is weak. In such a case, 
we can express conductance $G=dI_D/dV_D$ at $T=0$ as
\begin{equation}
G=g_0\frac{(E_F-\epsilon_f)^2}{(E_F-\epsilon_f)^2+z^2\eta^2}
\label{formula}
\end{equation}
This formula shows that $G$ has a dip structure when $\epsilon_f$
coincides with $E_F$.

\subsection{Two trap sites}
Here, we consider the two-trap-site case 
where two identical trap sites exist at $\pm {\bf R}/2$. 
As discussed in refs.~\cite{Newns}, \cite{Coleman}, and \cite{Tanamoto}, 
the symmetry of the two trap sites makes us 
set equal mean field values at $+ {\bf R}/2$ and $- {\bf R}/2$,  
such as $\epsilon_f^{(2)}\!\equiv\!\epsilon_f({\bf R}/2)\!=\!\epsilon_f(-{\bf R}/2)$
and $z\!\equiv\! z({\bf R}/2)=\!z(-{\bf R}/2)$. 
The mean field Hamiltonian for the two-impurity case, $
H_{\rm chan}^{\rm (II)}$, can be expressed
as a summation of two independent parts\cite{Newns,Coleman,Tanamoto}:
\begin{eqnarray}
\lefteqn{ H_{\rm chan}^{\rm (II)}=\!\!\sum_{ks}E_{k}c_{ks}^\dagger c_{ks} 
\!+\! \epsilon_f^{(2)} (n_{d1}\!+\!n_{d2}) }
\nonumber \\
&+&  \sqrt{z}\sum_{{\bf k}s}V_d 
[
c_{{\bf k}s}^\dagger \left(d_{1s}\exp{\left( i\frac{\bf k \cdot R}{2}\right)}\!
+\!d_{2s}\exp \left( {-i\frac{\bf k \cdot R}{2}}\right) \right)
\nonumber \\ 
\!&+&\! {\rm h.c.}] + 2(\epsilon_f^{(2)}\!-\!E_D)(z\!-\!1)
\nonumber \\
\!&=&\! \sum_{P=\pm} \{\sum_s 
E_k c_{{\bf k}s}^{P\dagger} c_{{\bf k}s}^P\!+\!\epsilon_f^{(2)} n^P_d
 \!+\!\sqrt{z} \sum_{{\bf k}s} \! V^P_d
 (c_{{\bf k}s}^{P\dagger} d_{s}^P\!+\! {\rm h.c.})\}\nonumber \\
&+&2(\epsilon_f^{(2)}\!-\!E_D)(z\!-\!1)
\end{eqnarray}
where $d_{1s}$ and $d_{2s}$ are respectively annihilation operators for the left (${\bf -R}/2$ )
and right (${\bf R}/2$) trap sites, 
$n_d^P\!=\!n_{d1}\!+P \!n_{d2} (n_{di}=d_{is}^\dagger d_{is})$, 
$d_s^P =(d_{1s}+P d_{2s})$, 
$V^P_d =V_d[2 N_P]^{1/2}$  
with $N_P \equiv (1+P \sin(k_FR)/k_FR)/2$ ($P=\pm$, $k_F$ is a wave vector 
at Fermi Energy), and 
\begin{eqnarray}
c_{ks}^{(+)}&=&\frac{1}{N_{+}} \int \frac{d\Omega_k}{4\pi}
\cos \left( \frac{\bf k \cdot R}{2} \right) c_{{\bf k}s}, \nonumber \\
c_{ks}^{(-)}&=&\frac{1}{N_{-}} \int \frac{d\Omega_k}{4\pi}
\sin \left( \frac{\bf k \cdot R}{2} \right) c_{{\bf k}s}.
\end{eqnarray}
Because the Hamiltonian is 
described by the two independent parts, 
$I_D$ and $G$ consist of two independent parts.
In particular, $G$ at $T=0$ is 
\begin{equation}
G=g_0\sum_{P=\pm}
\frac{(\epsilon_f-E_F)^2}
{(\epsilon_f-E_F)^2+z^2 \eta_P^2}
\label{eqn:cond_1b}
\end{equation}
where $\eta_P=\eta_{P0} y_0/(1+y_0)$ with 
$\eta_{P0}=(N_PV_d)^2 /\gamma$. 
Thus, the dip
in $G$ is intrinsic and can be described in a similar 
fashion to the single-trap-site case. 
\begin{figure}
\includegraphics[width=8.5cm]{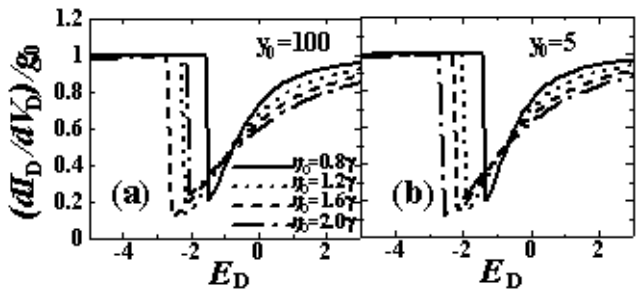}
\caption{Conductance $dI_D/dV_D$ for one trap site [Fig.1 (a)] 
as a function of trap site energy $E_D$ at $V_D=0.01\gamma$. 
(a) $y_0=100$. (b) $y_0=5$. $D=6\gamma$, $E_F=0$, and $T=0.01\gamma$.@
In this paper, we set $\Gamma_L=\Gamma_R$, 
and $\gamma$ as an energy unit.
}
\label{Fig.2}
\end{figure}

\section{Numerical Calculations}
Figure 2 shows the numerically calculated conductance $dI_D/dV_D$
at $V_D=0.01\gamma$ as a function of trap site energy $E_D$
when the coupling constant $\eta_0$ is changed.
We can see a deep dip structure near $E_F$.
This is the result of the interference between 
the channel electron and the trap site (FK effect) 
and shows that a trap site close to $E_F$ greatly 
degrades  device performance.
Figures 2(a) and 2(b) also show that the 
result is independent of the value of $y_0$, that is the DOS of 
the channel region.
Here, the minimum $T_K$ is larger than $T=0.01\gamma$. 
Figure 3 shows, when the dip appears, that the trap site is occupied 
by an electron ($z \sim 0$) and trap site energy $\epsilon_f$ 
increases.
\begin{figure}
\includegraphics[width=8.5cm]{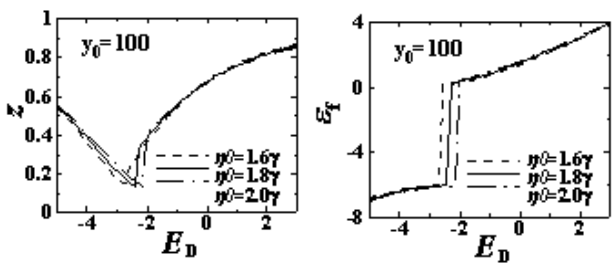}
\caption{Solutions of the self-consistent equations, eq.~(\ref{Self}).
(a) $z$ and (b) $\epsilon_f$ as a function of $E_D$. 
The parameters are the same as those in Fig.~2.
}
\label{Fig.3}
\end{figure}

Figure 4 shows the $I_D$-$V_D$ curve at $E_D=-1.2\gamma$ where the dip appears.
We can see that, as $\eta_0$ increases, $I_D$ decreases rapidly.  
This indicates that the existence of a trap site reduces  
drive current.
In Fig.~5, we calculate the conductance from eq.~(\ref{formula}) 
using $\epsilon_f$ and $z$ 
in Fig.~2. 
The conductance in Fig.~5 is almost the same as that in Fig.~2, 
and we found the simple formula eq.~(\ref{formula}),
in which a numerical integral is not required, 
to be very effective in most cases. 
\begin{figure}
\begin{center}
\includegraphics[width=5.0cm]{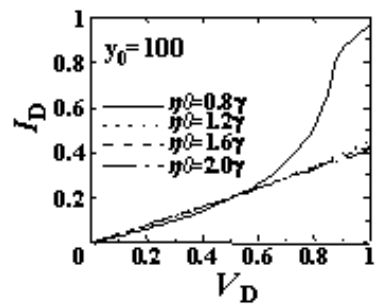}
\end{center}
\caption{$I_D-V_D$ characteristics for a trap site [Fig.~1(a)]. 
$E_D =-1.2\gamma$,  $y_0=100$, $D=6\gamma$, $E_F=0$, and 
$T=0.01\gamma$.}
\label{Fig.4}
\end{figure}

\section{Discussion}
Let us consider a simple estimation.
If we take $D\sim (\hbar^2/2m)(3\pi^2n)^{2/3}$ with 
an effective mass $m=0.2m_0$ ($m_0$ is the free electron 
mass) 
and a channel electron density $n\!=\!10^{17} $cm${}^{-3}$, 
then $D\sim$ 4 meV. 
Using $ N_c(E_F)=y_0/\gamma$ and $V_d^2 =\gamma \eta_0$, 
we have $T_K=D\exp (\pi E_D /(2y_0\eta))$.
If we assume $E_D\sim -\gamma$ and $\eta \sim \gamma$ in the 
above calculations, we have
$T_K\sim D\exp(-\pi/(2y_0))$. 
Thus, we obtain $T_K \sim 44.8$ K for $y_0=100$ and 
$T_K \sim 32.7$ K for $y_0=5$. 
These valuse of $T_K$ are much larger than that of 
an experiment of a QD, shown in ref.~\cite{Sato}. 
This is because we use large coupling constants
between the trap and the channel, and the confinement of 
the electron to the impurity trap site is considered to be stronger 
than that of the QD. 
More elaborate theoretical studies would be required 
to detect dip structures in ballistic transistors.
\begin{figure}
\begin{center}
\includegraphics[width=5.0cm]{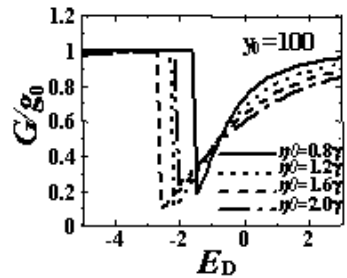}
\end{center}
\caption{Conductance formula eq.~(\ref{formula}) as a function of $E_D$, 
where self-consistent $z$ and $\epsilon_f$ are used. 
By comparing Fig.~5 with Fig.~2, we found that eq.~(\ref{formula}) is effective. }
\label{cond_formula}
\end{figure}

Changing $E_D$ corresponds to changing the gate bias 
if we assume that the gate bias dependence on $g_0$ and 
other quantities is sufficiently weak. 
In that case, it is expected that conductance has a dip structure when 
gate bias is changed. The measurement of gate bias dependence 
would be the easiest way to check the existence of 
the trap site and prove the FK effect. 

Here, we use an approximation 
in which quantities such as $N_c(E)$ are replaced 
by their values at $E_F$. This is because the FK effect is 
a result of the interference between a localized state and a continuum. 
To be more realistic, we should take into account 
the quantized vertical component and use a three-dimensional DOS, 
as in refs.\cite{Natori1} and \cite{Natori2}. 
A more precise relationship between charge the density $N_c(E)$ and 
the gate bias would be also required to apply our formula to actual devices. 

\section{Conclusions}
We studied the effect of trap sites on the transport 
of ballistic transistors, 
and showed that current is reduced and conductance 
has an intrinsic dip as a result of interference effect. 
This demonstrates an interesting interplay between physics 
and engineering devices.
We also derived an analytic form of conductance, 
which will help to analyze the existence of trap sites in experiments.

\acknowledgements{%
We thank N. Fukushima, A. Nishiyama, J. Koga, and R. Ohba for 
useful discussions.
}


\end{document}